\documentclass[a4paper,12pt]{article}
\usepackage{epsfig}
\usepackage{amssymb}
\usepackage{amsfonts}
\usepackage{amsmath}
\usepackage{euscript}
\usepackage{verbatim}
\usepackage{latexsym}
\usepackage{graphicx}

\newskip\humongous \humongous=0pt plus 1000pt minus 1000pt

\newif\ifdtup

\catcode`\@=11

\@addtoreset{equation}{section}

\def\@normalsize{\@setsize\normalsize{15pt}\xiipt\@xiipt
\abovedisplayskip 14pt plus3pt minus3pt%
\belowdisplayskip \abovedisplayskip
\abovedisplayshortskip \z@ plus3pt%
\belowdisplayshortskip 7pt plus3.5pt minus0pt}

\def\small{\@setsize\small{13.6pt}\xipt\@xipt
\abovedisplayskip 13pt plus3pt minus3pt%
\belowdisplayskip \abovedisplayskip
\abovedisplayshortskip \z@ plus3pt%
\belowdisplayshortskip 7pt plus3.5pt minus0pt
\def\@listi{\parsep 4.5pt plus 2pt minus 1pt
     \itemsep \parsep
     \topsep 9pt plus 3pt minus 3pt}}

\relax

\catcode`@=12

\catcode`\@=11

\def\section{\@startsection{section}{1}{\z@}{3.5ex plus 1ex minus
   .2ex}{2.3ex plus .2ex}{\large\bf}}

\def\SymBoxes#1#2#3#4{\newdimen\un@t \un@t#3%
\raisebox{#1}{\rule{#2\un@t}{#4}\hskip-#2\un@t
\@tempdimb\un@t \advance\@tempdimb by-#4\@tempcntb#2\relax%
\@whilenum{\@tempcntb>0}\do{
\rule{#4}{\un@t}\hskip\@tempdimb \advance\@tempcntb by\m@ne}%
\hskip-#2\un@t \rule[\un@t]{#2\un@t}{#4}%
\rule[\un@t]{#4}{#4}\hskip-#4
\rule{#4}{\un@t}}\hskip-#4}                

\begin{document}

\newcommand{\beq}{\begin{equation}}
\newcommand{\eeq}{\end{equation}}
\newcommand{\bea}{\begin{eqnarray}}
\newcommand{\eea}{\end{eqnarray}}
\newcommand{\beas}{\begin{eqnarray*}}
\newcommand{\eeas}{\end{eqnarray*}}
\newcommand{\defi}{\stackrel{\rm def}{=}}
\newcommand{\non}{\nonumber}
\newcommand{\bquo}{\begin{quote}}
\newcommand{\enqu}{\end{quote}}
\renewcommand{\(}{\begin{equation}}
\renewcommand{\)}{\end{equation}}
\def \eqn#1#2{\begin{equation}#2\label{#1}\end{equation}}
\def\IZ{{\mathbb Z}}
\def\IR{{\mathbb R}}
\def\IC{{\mathbb C}}
\def\IQ{{\mathbb Q}}
\def\de{\partial}
\def\Tr{ \hbox{\rm Tr}}
\def\H{ \hbox{\rm H}}
\def\HE{ \hbox{$\rm H^{even}$}}
\def\HO{ \hbox{$\rm H^{odd}$}}
\def\K{ \hbox{\rm K}}
\def\Im{ \hbox{\rm Im}}
\def\Ker{ \hbox{\rm Ker}}
\def\const{\hbox {\rm const.}}
\def\o{\over}
\def\im{\hbox{\rm Im}}
\def\re{\hbox{\rm Re}}
\def\bra{\langle}\def\ket{\rangle}
\def\Arg{\hbox {\rm Arg}}
\def\Re{\hbox {\rm Re}}
\def\Im{\hbox {\rm Im}}
\def\exo{\hbox {\rm exp}}
\def\diag{\hbox{\rm diag}}
\def\longvert{{\rule[-2mm]{0.1mm}{7mm}}\,}
\def\a{\alpha}
\def\dag{{}^{\dagger}}
\def\tq{{\widetilde q}}
\def\p{{}^{\prime}}
\def\W{W}
\def\N{{\cal N}}
\def\hsp{,\hspace{.7cm}}

\def\br{\nonumber\\}
\def\IZ{{\mathbb Z}}
\def\IR{{\mathbb R}}
\def\IC{{\mathbb C}}
\def\IQ{{\mathbb Q}}
\def\IP{{\mathbb P}}
\def \eqn#1#2{\begin{equation}#2\label{#1}\end{equation}}

\newcommand{\C}{\ensuremath{\mathbb C}}
\newcommand{\Z}{\ensuremath{\mathbb Z}}
\newcommand{\R}{\ensuremath{\mathbb R}}
\newcommand{\rp}{\ensuremath{\mathbb {RP}}}
\newcommand{\cp}{\ensuremath{\mathbb {CP}}}
\newcommand{\vac}{\ensuremath{|0\rangle}}
\newcommand{\vact}{\ensuremath{|00\rangle}                    }
\newcommand{\oc}{\ensuremath{\overline{c}}}
\begin{titlepage}
\begin{flushright}
CHEP XXXXX
\end{flushright}
\bigskip
\def\thefootnote{\fnsymbol{footnote}}

\begin{center}
{\Large
{\bf Subttractors
}
}
\end{center}

\bigskip
\begin{center}
{\large  Avik Chakraborty$^a$\footnote{\texttt{avikchakraborty88@yahoo.in}} and
Chethan Krishnan$^a$\footnote{{\texttt{chethan@cts.iisc.ernet.in}}}}
\vspace{0.1in}

\end{center}

\renewcommand{\thefootnote}{\arabic{footnote}}

\begin{center}
$^a$ {Center for High Energy Physics\\
Indian Institute of Science, Bangalore, India}\\

\end{center}

\noindent
\begin{center} {\bf Abstract} \end{center}
We consider extremal limits of the recently constructed ``subtracted geometry". We show that extremality makes the horizon attractive against scalar perturbations, but radial evolution of such perturbations changes the asymptotics: from a conical-box to flat Minkowski. Thus these are black holes that retain their near-horizon geometry under perturbations that drastically change their asymptotics. We also show that this extremal subtracted solution (``subttractor") can arise as a boundary of the basin of attraction for flat space attractors. We demonstrate this by using a fairly minimal action (that has connections with STU model) where the equations of motion are integrable and we are able to find analytic solutions that capture the flow from the horizon to the asymptotic region. The subttractor is a boundary between two qualitatively different flows. We expect that these results have generalizations for other theories with charged dilatonic black holes.

\vspace{1.6 cm}
\vfill

\end{titlepage}

\setcounter{footnote}{0}

\section{Introduction and Conclusion}
\label{intro}

Most of the progress in understanding black holes in string theory has been in the context of supersymmetric black holes \cite{StromingerVafa}. Many such black holes exhibit a phenomenon known as the attractor mechanism \cite{FKS}, which is the statement that the moduli scalars that are present in the geometry can exhibit non-trivial radial profiles even when their values are fixed at the horizon. This is interesting for many reasons. One reason is that the horizon values of the scalars are determined by the charges of the black hole\footnote{This is a generic statement. The scalars that have no coupling to the gauge fields (or curvature) are not fixed by the gauge charges \cite{SenReview}.}, which in turn are responsible for its thermodynamic properties. Therefore, the attractor mechanism is an indication that a microscopic understanding of black holes is likely to require only the near-horizon data\footnote{This statement has exceptions. See eg. \cite{SenException}.}. 
This intuition is consistent with AdS/CFT \cite{Maldacena} and holography \cite{tHooft, Susskind}. Another reason that makes the attractor mechanism interesting is that the asymptotic values of the scalars have a coupling constant interpretation, and therefore one can tune the scalar value at infinity to weak coupling and hope to understand an otherwise intractable problem \cite{DabholkarSen}. 

Supersymmetric black holes are necessarily extremal, but not vice versa. Interestingly, it was shown by Goldstein et al. \cite{Goldstein} that the attractor mechanism relied only on the extremality of the black hole and not its supersymmetry. In particular, they showed that when the black hole is extremal, perturbing the moduli scalars near the (regular, double-zero) horizon  results in solutions that change the asymptotic values of these scalars. This demonstrates the attractor mechanism. They showed that this phenomenon was robust for flat and anti-de Sitter boundary conditions. The perturbation of the scalar retained the asymptotic geometry, even though it moved the scalar value at infinity. One of our goals in this paper is to show that there exist fully controllable solutions of the Einstein-Maxwell-dilaton system which correspond to extremal black holes such that perturbations change the asymptotic structure completely, while preserving the near-horizon geometry.

One piece of folk lore regarding black holes in general, and not just extremal black holes, is that they should have a (holographic) description in terms of a conformal field theory (CFT). A crude way to motivate this is to note that an observer at infinity sees any object falling into the black hole as getting infinitely redshifted as it gets close to the horizon, and therefore a putative holographic dual of the spacetime might correspond to an IR fixed point CFT. A less philosophical, more concrete argument is that a very large class of (non-extremal) black holes in string theory have formulas for their thermodynamic quantities that look tantalizingly like those of a CFT \cite{CveticLarsenOld}. There has been some recent progress in making these ideas more concrete, starting with the Kerr-CFT correspondence \cite{StromKerrCFT} and its generalizations \cite{LMP, Hartman, Stanislav, Geoffrey}. It was realized that even non-extremal Kerr black holes exhibit a hidden conformal symmetry at the level of the wave equations that propagate on them \cite{HiddenCS}. Following up on this work, it was shown in \cite{CK} that such a hidden conformal symmetry can also be seen in the very general black hole geometries of \cite{CveticLarsenOld}. 

In some recent work, Cvetic and Larsen have shown \cite{CL1, CL2} (see also Cvetic and Gibbons \cite{CG}\footnote{See \cite{CG, Ami} for a Harrison transformation approach to the subtracted geometry.}) that this near-region hidden conformal symmetry at the level of the wave equations can be implemented at the level of the geometry, if one replaces a single function (``warp factor") in the original black hole geometry with a new function (which can be determined in terms of the parameters of the original black hole). Wave equations in this ``subtracted geometry" are by construction in the (effective) BTZ form. Interestingly, the replacement of the warp factor retains the black hole interpretation and does not affect the thermodynamic quantities of the black hole, but it now lives in an asymptotically conical box \cite{CG} instead of the original asymptotically flat space. To support the geometry, the matter content has to be appropriately modified. In \cite{CG}, static black holes in four-dimensional ${\cal N}=2$ supergravity coupled to four vector multiplets were used as the theory in which such ``subtracted geometries" can be studied. We can work with three scalar fields and four vector fields in the bosonic sector, when we restrict to static spacetimes. This action is quite general and allows black holes with four distinct charges. Both the original black hole and its subtracted geometry can be found as solutions of this system. 

In this paper, we will mostly be concerned with extremal limits of static charged black holes and we will write down their corresponding subtracted geometries. We consider a fairly minimal theory (engineered from ${\cal N}=2$ supergravity coupled to four vector multiplets), where these solutions can be studied. Since it is known that extremal black holes are attractive, we will investigate the nature of the attractor mechanism in the context of the subtracted metric. We observe that the scalars and the $U(1)$ gauge fields that support the subtracted geometry fall into the attractor ansatz presented in \cite{Goldstein}. To study the attractor mechanism we perturb the scalar near the horizon (while demanding regularity at the double zero horizon) and trace the radial evolution of the system. Numerically, we find that the asymptotic structure changes drastically, and that the solution flows to an asymptotically {\em flat} (extremal) solution. To get more confidence in our result, we also investigate the system {\em analytically}. For our theory, using a connection to an integrable Toda system \cite{GibMaed, Pope, Goldstein} we solve the EoMs exactly. We show that the extremal subtracted geometry arises as a (previously missed) solution of this system, and show analytically that perturbations of this geometry flow to flat space extremal black holes with non-trivial radial scalar profiles. The entire discussion here is under analytic control and we find a consistent picture. The conclusion is fully in accord with our numerical experiments and we conclude that the extremal subtracted geometry (the ``subttractor") can retain its near-horizon features, in particular the scalar values and the charge parameters, even though the asymptotic structure changes drastically. We also observe that these subttractors can be interpreted as a boundary of the domains of attraction of flat space attractors. In fact when we perturb our subttractors one way, they flow to flat space hairy black holes, but when we perturb them the other way we run into new exact solutions with no asymptotic region. The phase structure of attractor solutions is quite rich, and we have evidence that a generalized notion of subtracted geometry might be useful for understanding them. But we leave a detailed discussion of these matters for a future paper \cite{K}.

The outline of this paper is as follows. In the next section, we introduce a fairly minimalist action where subtracted geometries can be found. We present the static charged black hole solutions, their subtracted geometries and the corresponding extremal limits. In Section \ref{AttractAnsatz}, we present the attractor ansatz and the equations of motion and observe that the subtracted geometry falls into that ansatz. In Section \ref{Exact}, we use the connection of our system to a Toda-style system to construct various exact solutions. In Section \ref{subtt}, we do perturbation theory around the horizon and identify the solutions to which the subttractor flows when perturbed. We do this by finding the precise relation between the asymptotic value of the scalar and the perturbation at the horizon. We conclude by showing that our solutions arise at the boundary of attraction basin in asymptotically flat attractors. In an appendix we present a rationale for obtaining the symplified (integrable) action that we use, from ${\cal N}=2$ supergravity in four dimensions coupled to four vector multiplets. In another appendix we present some comments about asymptotically conical box geometries.

\section{A Minimalist Subtracted Geometry}
\label{SubEx}

We will be interested in black hole solutions of the action
\bea \label{minaction}
S=\int d^4x \Big(R-2 (\partial \phi)^2- f_{ab}(\phi)F^{a}_{\mu\nu}F^{b \ \mu\nu} \Big)
\eea
where the scalar-dependent gauge couplings take the form
\bea
f_{ab}(\phi)=\left(
\begin{array}{cc}
e^{\alpha_1 \phi}& 0 \\
0& e^{\alpha_2 \phi}
\end{array}
\right),
\eea
where $(\alpha_1, \alpha_2)=(2\sqrt{3}, -2/\sqrt{3})$. This theory has one dilatonic modulus scalar and two $U(1)$ gauge fields, and the $f_{ab}$ is the gauge coupling matrix.

This action contains all the low spin (0, 1 and 2) bosons with long range classical effects and there are a few reasons why we have chosen this specific form. The scalar does not have a potential, i.e., its a modulus, and the gauge-coupling is scalar dependent. Both these features are natural in toroidal Kaluza-Klein reduction scenarios. Specifically, we show in an appendix that this action can be obtained from the bosonic sector of ${\cal N}=2$ supergravity coupled to four vector multiplets, which is closely related to the context in which Cvetic and Youm \cite{CveticYoum} wrote down their very general 5D black holes in string theory. The subtracted geometry for the four-dimensional black holes was presented in \cite{CL2}.

Our action has the advanatge that it allows analytically tractable static black hole solutions with non-trivial scalar profiles \cite{GibMaed, Pope, Goldstein}. Specifically, it is known that explicit attractor solutions can be written down \cite{Goldstein}.

A final reason why we work with this action as opposed to a more general action like the STU model is that this is a fairly {\em minimal} set up where one can write down a subtracted geometry that is a solution of the same system of equations of motion as the original black hole\footnote{We only consider static black holes in this paper.}. In more general theories, like the ones considered in \cite{CL2, CG}, the subtracted metric has only 3 parameters wheras the original geometry has more. This means that many black holes have the same subtracted geometry in the construction of \cite{CL2, CG}. Conversely, if we work with less complicated theories (like say pure gravity) the subtracted geometry cannot be thought of as the solution of the same action.

Using the connection with the action of \cite{CL2} discussed in the appendix, it is straightforward to write down static black hole solutions of this action with two $U(1)$ charges. The metric takes the form
\bea\label{metric}
ds^2=-\frac{X}{\sqrt{\Delta_0}} dt^2+\frac{\sqrt{\Delta_0}}{X}dr^2+\sqrt{\Delta_0}\ (d\theta^2+\sin^2 \theta d\varphi^2)
\eea
where 
\bea
X=r(r-2m), \ \Delta_0=(r+2m \sinh^2 \delta)^3 (r+2m \sinh^2 \delta_0)
\eea
with the matter fields given by
\bea
\exp \frac{2\phi}{\sqrt{3}}=\Big[\frac{r + 2 m \sinh^2
\delta}{r + 2 m \sinh^2 \delta_0}\Big]^{1/2}, \hspace{0.5in}\\
F^1= Q^1_m\sin \theta \ d\theta \wedge d\phi, \ \ \ F^2= Q^2_m  \sin \theta \ d\theta \wedge d\phi,
\eea
where we have taken a magnetic form for the gauge fields with magnetic charges given by
\bea
Q^1_m= \frac{m \sinh 2\delta_0}{2}, \ \ \ Q^2_m= \frac{\sqrt{3} m \sinh 2\delta}{2}. \label{magcharge}
\eea

We will also write down the subtracted geometry for this solution. The metric takes the same structure as above, but now the warp factor $\Delta_0$ is replaced by
\bea\label{subrep}
\Delta_s=(2m)^3\ r \ (\Pi_c^2-\Pi_s^2)+(2m)^4\ \Pi_s^2
\eea
where 
\bea\label{pipi}
\Pi_c=\cosh^3 \delta \ \cosh \delta_0, \ \ \ \Pi_s =\sinh^3 \delta \sinh \delta_0
\eea
and the matter fields take the form
\bea
\exp \frac{2\phi}{\sqrt{3}}=\frac{\beta^2}{\sqrt{\Delta_s}},
\eea
and the magnetic fields and the charges take the same form as in the unsubtracted case. 

The value of $\beta^2$ in terms of $m, \delta$ and $\delta_0$ is presented in Section \ref{Exact}. In principle, we can choose to rescale the charges (both in the subtracted and unsubtracted geometry) if we do a compensating shift in the scalar fields, which corresponds to tuning $\beta$. This is clear from the structure of our action, and will also be evident when we write down the equations of motion for the system using an effective potential in the next section along the lines of \cite{Goldstein}. But we want to work with the same specific set of magnetic charges (\ref{magcharge}) for both the sub(and unsub-)tracted geometries, so this means that we have eliminated this freedom. In particular, the equations of motion deteremine the $\beta^2$ for us. This is different from the situation considered in \cite{CG} where they left this scaling essentially arbitrary. We think it is more interesting to compare the two geometries when their charges are identical, and the results of this paper give evidence for this. But we emphasize that if one's goal is only to find {\em some} matter to support the subtracted metric, then the system has more redundancies. This is even more apparent when we turn on more charges in the original solution \cite{CL2, CG}, because then there are many unsubtracted metrics which correspond to one subtracted metric. But choosing the charges fixes a one-to-one map between the original black hole and its subtracted solution. Note however that this prescription only works for black holes with charge - in particular, the Schwarzschild black hole is outside our jurisdiction. This won't bother us however, since we are ultimately interested in the attractor mechanism which is a statement about fixed non-zero charges, extremal black holes, etc. The subtracted Schwarzschild black hole can be seen as a degenerate limit of our set up, where the $\beta^2$ along with the magnetic charges goes to zero. In particular, this means that the scalar has a constant piece that diverges ($\rightarrow -\infty$). Because of this the effective potential (\ref{mineff}) that we will define in the next section will diverge for subtracted Schwarzschild. 

The metric with $\Delta_s$ as the warp factor is not asymptotically flat. Indeed, it is straightforward to show that the geometry asymptotes to a conical box. Some basic properties of this kind of asymptotic geometry are discussed in an appendix. 

The black holes we considered have extremal limits. Specifically, we will look at the extremal cases corresponding to the limit
\bea
\label{extlim}
m \rightarrow 0, \ \ \ \delta_0, \delta \rightarrow \infty
\eea
such that the charges $Q^1_m$ and $Q^2_m$ are held fixed. This is sometimes called the ``BPS limit" when working in the context of the STU model. The mass of the black hole in terms of the parameters $m, \delta$ and $\delta_0$ is given by
\bea
M=\frac{m}{4}(3 \cosh 2 \delta + \cosh 2 \delta_0) 
\eea
and therefore in the extremal limit, we have the relation 
\bea
M=\frac{Q^1_m}{2}+\frac{\sqrt{3}\ Q^2_m}{2}.
\eea
The subtracted geometry takes a simple form in this limit, and is defined by
\bea
X= r^2, \ \ \ \Delta_s=\frac{\alpha^4}{\mu} (r +\mu)
\eea
where
\bea \label{subtcharge}
\Big(\frac{\alpha}{4}\Big)^4\equiv \Big(\frac{Q^2_m}{2\sqrt{3}}\Big)^3\ \frac{Q^1_m}{2}, \ \ \ \ \frac{4}{\mu}\equiv\frac{6 \sqrt{3}}{Q^2_m}+\frac{2}{Q^1_m}
\eea
In this coordinate system, the (double-zero) horizon is at $r=0$. There are only two independent parameters that define the metric because the mass is determined in terms of the two charges in the extremal limit. The solution is fully determined once we give $\beta^2$, which is given in Section \ref{Exact}.

\section{The Attractor Ansatz} 
\label{AttractAnsatz}

The extremal subtracted geometry is what we call the subttractor. The reason for this is that when we allow perturbations of this geometry close to the horizon that are finite at the horizon, the horizon value of the scalar does not change. This is one way in which an attractor black hole can be characterized \cite{Goldstein}. 

The basic observation that makes the connection with the attractor mechanism possible is that both the original black hole and the subtracted geometry fall into what we call the ``attractor ansatz"\cite{Goldstein}. For our purposes, this means that the fields are assumed to have only radial dependence and that they are of the form
\bea
ds^2=-a(r)^2 dt^2+\frac{dr^2}{a(r)^2}+b(r)^2 d \Omega^2, \\
F^{a=1,2}=Q^{a=1,2}_m\sin \theta d\theta \wedge d\phi, \ \ \ \phi \equiv \phi(r) \hspace{ 0.1in}
\eea
and therefore the equations of motion can be brought to a simple form. Clearly, the black holes we have discussed so far all fall into this ansatz. Note in particular that the attractor ansatz is {\em not} limited to extremal black holes. Even though we won't be using it directly in this paper, it is also true that the solutions considered in \cite{CL2, CG} all fall into the attractor ansatz. We will return to them elsewhere \cite{AAC}.

The equations of motion take a simple and useful form-
\bea\label{minansatze1}
(a^2\ b^2)''-2 =0 \\
\frac{b''}{b}+{\phi '}^2=0 \\
(a^2 b^2\phi')'-\frac{\partial_\phi V_{eff}(\phi)}{2 b^2}=0
\eea
together with a (redundant) first order energy constraint
\bea 
a^2{b'}^2+{a^2}'{b^2}'+\frac{V_{eff}(\phi)}{b^2}-a^2b^2 {\phi'}^2-1=0,\label{minansatze2}
\eea
where 
\bea\label{mineff}
V_{eff}(\phi)=e^{\alpha_1 \phi} (Q^1_m)^2+e^{\alpha_2 \phi} (Q^2_m)^2
\eea
As we discussed in the previous section, we can shift the scalars by  a constant, while rescaling the charges by a compensating factor, and the equations of motion will remain invariant.

For the specific action that we have chosen, these equations can (perhaps surprisingly) be exactly integrated, by relating it to a Toda-like system \cite{GibMaed, Pope, Goldstein}. The integrability/diagonalizability has some further generalizations and we will investigate some of these questions in \cite{AAC}. We will show that the subttractor geometry that we have presented in the last section can be found as a previously missed solution of this Toda-like system. The advantage of the exact solution is that we can perturb the subttractor geometry near the horizon, and trace the evolution of the system analytically. This is what we set out to do in Section \ref{subtt}. We present the various exact solutions relevant to us in the next section. 

\section{Exact Solutions}
\label{Exact}

Here, following \cite{Goldstein} we present the exact solutions of the equations of motion in Section \ref{AttractAnsatz}. There are two reasons we do this. Firstly, we want to identify the subtracted geometry and the subttractor among the possible solutions of this system. The extremal flat (attractor) solutions are discussed in \cite{Goldstein}. We repeat them here for the slightly more general non-extremal case, do the matches with the black hole solutions we presented and determine the integration constants.  Our expressions in this case turn into those presented in \cite{Goldstein} when we go to the extremal limit. Next, we also show how one can see the subtracted geometry and its extremal limit (subttractor) using these equations. 

The general solution \cite{GibMaed, Pope, Goldstein}, after demanding regularity at the outer horizon gives us 
\bea
  \label{generalform1}
  e^{(\alpha_1-\alpha_2)\phi}&=&\displaystyle
    \left(
      -\frac{\alpha_2}{\alpha_1}
    \right) 
    \left(
      \frac{Q^2_m \sinh \ m (\tau-d_2)}{Q^1_m \sinh \ m (\tau-d_1) }
    \right)^2 \\
  a^2 &=& \displaystyle
     \left(\frac{ m}{Q^1_m \sinh \ m (\tau-d_1)}\right)^{\frac{-2\alpha_2}{\alpha_1-\alpha_2}} 
      \left(
        \frac{ m}{Q^2_m \sinh \ m (\tau-d_2)}
      \right)^{\frac{2\alpha_1}{\alpha_1-\alpha_2}}
     \\
  b^2 &=& r(r-2m)/a^2\label{generalform2}
\eea
where we have chosen coordinates so that the inner horizon is at $r=0$ and we have denoted the location of the outer horizon to be at $r=2m$. The tortoise coordinate
\bea
\tau=\frac{1}{2m}\log \Big(1-\frac{2m}{r}\Big).
\eea
This matches our conventions for the black hole solutions in Section \ref{SubEx}. These are the most general solutions once we demand regularity at the horizon, and they have a good limit in the extremal limit, namely when $m\rightarrow 0$. There are two remaining integration constants, $d_1$ and $d_2$. If we are interested in flat space solutions, they can be fixed by demanding (a) asymptotic flatness, i.e, $a(r \rightarrow \infty)=1$ and (b) choosing the asymptotic value of the scalar, i.e., $\phi(r \rightarrow \infty)=\phi_\infty$. 

These conditions fix $d_1$ and $d_2$ via
\bea
\sinh ( m \ d1)&=& \frac{m}{2\ Q^m_1}\exp({-\sqrt{3}\ \phi_\infty}), \\
\sinh ( m \ d2)&=&  \frac{m\sqrt{3}}{2\ Q^m_2}\exp\Big(\frac{{\phi_\infty}}{\sqrt{3}}\Big).
\eea
We have checked that these solutions match the flat space (non-extremal, unsubtracted) black holes we wrote down in section \ref{SubEx} when we set $\phi_\infty=0$. The solutions with non-zero $\phi_\infty$ go to the hairy extremal solutions investigated in \cite{Goldstein}\footnote{modulo some inconsequential but frustrating typos in \cite{Goldstein}. The solution is rather complicated, but we hope and believe that we haven't introduced our own typos in the process of correcting theirs. Sigh.} when we send $m \rightarrow 0$. When we perturb our subttractors they flow to these solutions, modulo one generalization- the asymptotic value of $a(r)$ need not be unity, it can be a constant $a_0$. There is a scaling symmetry in the system of equations (\ref{minansatze1}-\ref{minansatze2})
\bea\label{rscaling}
a\rightarrow t \ a, \ \ \ r \rightarrow t \ r 
\eea
for arbitrary $t$ so this $a_0$ can be gotten rid of by scaling the $r$. In any event, we present these flat space solutions, including the $a_0$, here in our notation:
\bea\label{hairy1}
\phi_{\rm hairy}(r)&=&\frac{\sqrt{3}}{4}
  \log\Big[\frac{
    2\ Q^2_m + 
     \sqrt{3}\  e^{\phi_\infty/\sqrt{3}}\  r/a_0}{
  \sqrt{3} \ (2  \ Q^1_m + \ e^{-\sqrt{3} \phi_\infty}r/a_0 )}\Big] 
  \\
b_{\rm hairy}(r)&=&3^{-3/8}\Big(2\ Q^2_m + 
     \frac{r \sqrt{3}\  e^{\phi_\infty/\sqrt{3}}}{a_0} \Big)^{3/4}\Big(2  \ Q^1_m + \frac{r\ e^{-\sqrt{3} \phi_\infty}}{a_0} \Big)^{1/4}, \\ 
     a_{\rm hairy}(r)&=&r/b_{\rm hairy}(r).\label{hairy2}
\eea

One crucial observation is that we can bring the general solutions (\ref{generalform1}-\ref{generalform2}) to the form of the subtracted geometry of section \ref{SubEx}, by setting $d_2=0$ and choosing $d_1$ appropriately. Setting $d_1=0$ and retaining $d_2$ results in a new type of solution which is neither the subtracted nor flat geometry: in fact, the warp factor $\Delta$ goes as $r^3$ in these solutions as $r \rightarrow \infty$, whereas it goes like  $r^4$ in flat black holes and like $r$ in the subtracted geometry. These new geometries also seem to have some interesting properties, and will be discussed elsewhere \cite{K}. 

We can match the subtracted solutions with the  $d_2=0$ solutions by taking 
\bea\label{d1unsub}
d_1=\frac{1}{m}\sinh^{-1} \Big(\frac{
 15 \cosh 2\delta + \cosh 6 \delta + 
  2 (5 + 3 \cosh 4 \delta) \cosh 2 \delta_0}{4 \sinh^3 2 \delta \sinh 2 \delta_0}\Big).
\eea 
The value of $\beta^2$ in the subtracted geometry that we presented in section \ref{SubEx} is given by
\bea
\beta^2=2 \sqrt{2}\ m^2\sqrt{\frac{\sinh 2\delta\ (\cosh^6 \delta \cosh^2 \delta_0 - 
  \sinh^6 \delta \sinh^2 \delta_0)}{ \sinh 2 \delta_0 \sinh\  m d_1}}
\eea
We have not been able to simplify these expressions in a useful way (if at all it is possible), but we have checked that they reproduce both the exact Toda-style solution (with the subtracted geometry parameters), as well as the subtracted geometry presented in section \ref{SubEx}. A more useful expression than the explicit form of $\beta^2$ is its value in the extremal limit. This can be found via the limit (\ref{extlim}) and the result, when written in terms of the gauge charges that support the geometry takes the simple form
\bea
\beta^2_{\rm Ext}=\frac{4 (Q^2_m)^2}{3}.
\eea
This fixes the horizon value of the dilaton to be the attractor value, both in the subtracted and unsubtracted geometries (in the extremal limit) to be
\bea
 \phi_{\rm Hor}=\frac{\sqrt{3}}{4}\log\Big(\frac{Q^2_m}{\sqrt{3}\  Q^1_m}\Big).
\eea

Now we present the extremal limit subtracted solutions (``subttractor") in a form that we have used for horizon perturbation computations.
\bea
\phi_s(r)=\frac{\sqrt{3}}{4} \log\Big(\frac{Q^2_m}{\sqrt{3} (Q^1_m + d_1 Q^1_m r)}\Big), \hspace{0.5in} \\
b_s(r)=2\Big(\frac{(Q^2_m)^3\ (Q^1_m + d_1 Q^1_m r)}{3 \sqrt{3}}\Big)^{1/4}, \ \ a_s(r)=r/b_s(r).
\eea
The comparison with the extremal subtracted solutions we presented towards the end of section \ref{SubEx} is established when we identify
\bea
d_1=1/\mu.
\eea
It can be checked that the complicated expression for $d_1$ in the non-extremal case (\ref{d1unsub}) reduces to this when we go to the extremal limit, when we use the relations (\ref{subtcharge}). For the extremal subtracted ``subttractor" solutions we presented in Section \ref{SubEx}, $\mu$ is not arbitrary like in the solution above - it is fixed in terms of the charges via (\ref{subtcharge}). This can again be understood in terms of the rescaling symmetry (\ref{rscaling}).

\section{Subttractor}
\label{subtt}

Our goal in this section is to start with the subttractor geometry, perturb it around the horizon and see what it evolves to in the asymptotic region. We will find that the solutions flow to flat space black holes with non-trivial radial dilaton profile. We will work with the exact form of the solution presented in Section \ref{Exact}. 

The idea is to perturb the scalar near the horizon and look for solutions of the equations of motion:
\bea
\phi(r)&=&\phi_s(r)+\delta \phi(r), \\ 
b(r)&=&b_s(r)+\delta b(r), \\ 
a(r)&=&a_s(r)+\delta a(r),
\eea
where $\phi_s(r), b_s(r), a_s(r)$ correspond to the subttractor solution that we have described in the appendix. Since these are second order equations, there will be two solutions (for each function) and we will consider the solution that stays finite {\em at} the horizon so that the double-zero structure of the horizon is maintained \cite{Goldstein}. In other words, we demand that the perturbation be regular at the horizon. Demanding that the equations of motion are satisfied order by order in $r$ (around the horizon $r=0$) then fixes the perturbed solution order by order in terms of one integration constant, which we take to be the ${\cal O}(r)$ piece in $\delta \phi(r)$. We will call this constant $\phi_{1}$. Concretely, when we do all this, we find that
\bea \label{horpert1}
\phi(r)=\frac{\sqrt{3}}{4}\log\Big(\frac{Q^2_m}{\sqrt{3}\  Q^1_m}\Big)+r \Big(\phi_1-\frac{\sqrt{3}}{4\mu} \Big)+r^2 \Big(\frac{\sqrt{3} }{8 \mu^2} - \frac{3 \phi_1}{4 \mu}  + 4  \frac{\phi_1^2}{\sqrt{3}} )\Big)+ \dots, \\ 
b(r)=\alpha\ \Big[1 +  \frac{r}{4\mu} + 
\frac{1}{32} \Big( \frac{8 \sqrt{3} \phi_1}{\mu}-\frac{3}{\mu^2} 
- 16 \phi_1^2\Big)\  r^2+\dots\Big], \hspace{0.6in} \\ 
a(r)=\frac{1}{\alpha}\Big[r - \frac{r^2}{4\mu} + 
 \frac{1}{32}\Big(\frac{5}{\mu^2} -\frac{8 \sqrt{3} \phi_1}{\mu} + 16 \phi_1^2\Big) r^3+\dots \Big]. \hspace{0.65in}\label{horpert2}
\eea 
When $\phi_1$ is set to zero this reduces to the subttractor solution $\phi_s(r), b_s(r), a_s(r)$, series expanded in $r$ around the horizon. We have defined $\alpha$ and $\mu$ in (\ref{subtcharge}). Note that the perturbation vanishes at the horizon, which means that the attractor mechanism is at work at the double-zero horizon. 

The possibility of the simple power series expansion we have used here is crucially tied to the choice of parameters $\alpha_1=2 \sqrt{3}, \ \alpha_2=-2/\sqrt{3}$, that we made in the dilatonic gauge-coupling matrix in Section \ref{SubEx}. For more general values of the couplings, we will typically expect to have irrational exponents in the series expansion above. This is tied to the fact that the near-horizon extremal geometry has an $AdS$ structure and the exponents are determined by the solutions of scalar wave equation in that geometry. This in turn depends on the effective cosmological constant of this near-horizon $AdS$, which depends indirectly on the dilatonic gauge couplings. One reason we have chosen to look at this specifically engineered theory is that the relevant exponent $\gamma$ turns out to be \cite{Goldstein}
\bea
\gamma=\frac{1}{2}(\sqrt{1-2 \alpha_1 \alpha_2} -1),
\eea
which reduces to unity in our case. This makes the power series expansion significantly simpler to work with in Mathematica, even though the attractor mechanism is qualitatively unaffected by this simplification. Ultimately, this simplification is to be traced to a connection to the attractor ansatz in ${\cal N}=2$ supergravity that we discuss in the appendix.

What we would like to see now, is how one can evolve the perturbed solution radially outward once we choose $\phi_1$. This is straightforward to do numerically for specific values of $Q^1_m, Q^2_m$ and $\phi_1$. Indeed, we have done this and checked that the solution goes to an asymptotically flat spacetime. But we can do more. Since we have shown in Section \ref{Exact} that solutions can be found exactly, and since we have found numerically that the solution goes to flat space, we can hope to relate the perturbation $\phi_1$ to the parameters of an asymptotically flat (extremal) solution. This is easy enough to do by taking the exact flat space solution, expanding it around the horizon and comparing it against the perturbed solution around the subttractor. We can relate $Q^1_m, Q^2_m$ and $\phi_1$ in the perturbed subttractor solution to $Q^1_m, Q^2_m$ and $\phi_\infty$ in the flat space hairy solution by comparing low order terms. Once this is done, all the higher order terms should match automatically if the solution is flowing to asymptotically flat space. This is indeed what we find. 

More concretely, the horizon expansion of the flat space hairy solution presented in (\ref{hairy1}-\ref{hairy2}) matches (\ref{horpert1}-\ref{horpert2}) when we set
\bea \label{check1}
\phi_1&=&\frac{3 \exp \Big(\phi_\infty/\sqrt{3}\Big)}{2 a_0\ Q^2_m}, \\
\frac{1}{\mu}&=&\frac{\exp (-\sqrt{3}\ \phi_\infty)}{2 a_0\ Q^1_m}+\frac{3 \sqrt{3}\ \exp \Big(\phi_\infty/\sqrt{3}\Big)}{2 a_0 \ Q^2_m} \label{check2}
\eea
Note that $\mu$ can be thought of as arising from a freedom to do scaling \`a la (\ref{rscaling}) and $a_0$ captures that.
 
\begin{figure}
\begin{center}
\includegraphics[height=0.35\textheight]{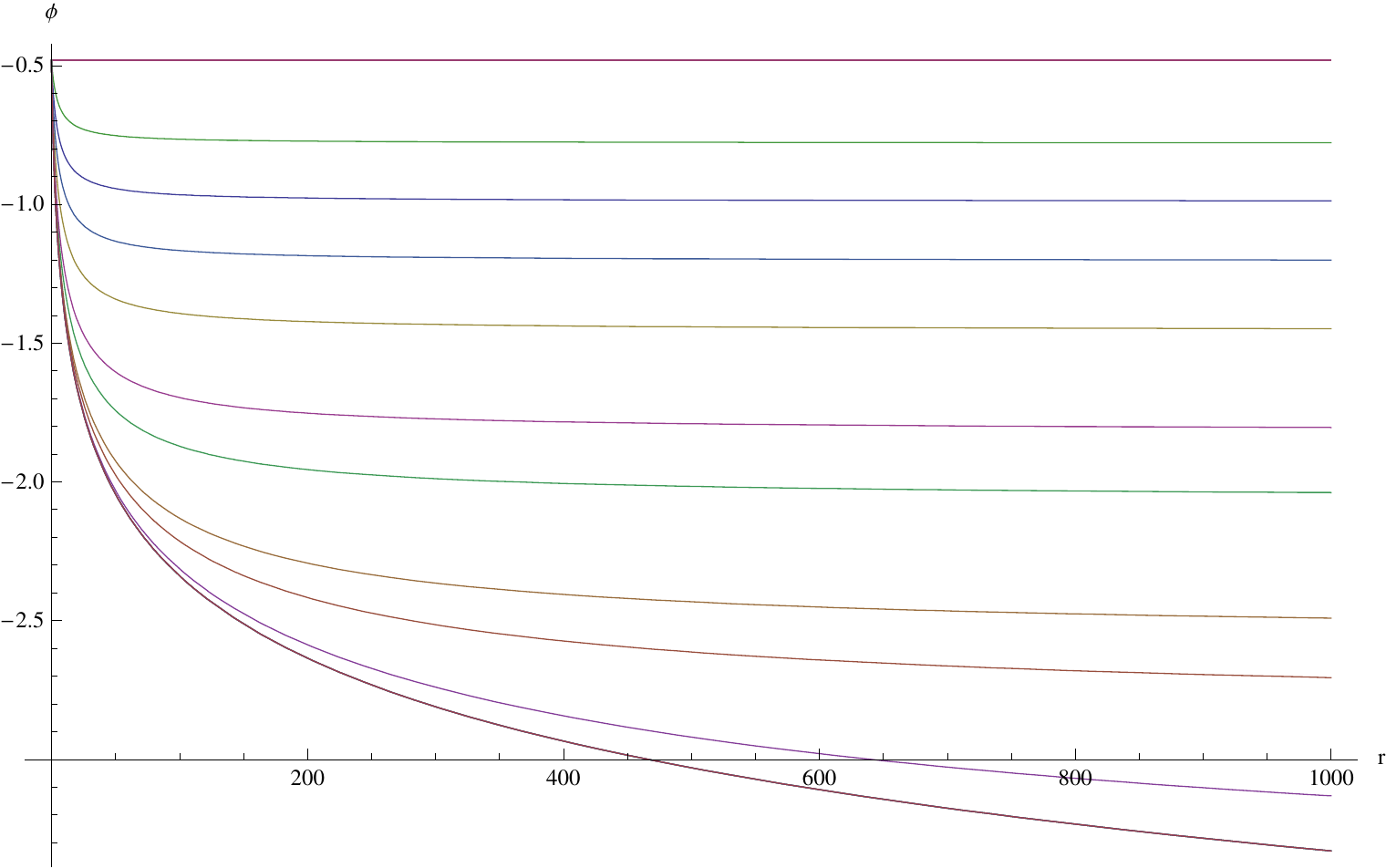}
\caption{Evolution of subttractor perturbations for $Q^1_m=7,Q^2_m=4$. The lowest curve is the subttractor, which runs off to infinity logarithmically. The perturbed solutions all stabilize to some finite $\phi_\infty$. The horizon value of the scalar is $\frac{\sqrt{3}}{4}\log\Big(\frac{Q^2_m}{\sqrt{3}\ Q^1_m}\Big)=-0.480177$.
}
\end{center}
\end{figure}

We plot the solutions for a specific choice of $Q^1_m,\ Q^2_m$ in figure 1. We have checked for all the cases that the asymptotic values of these solutions match with $\phi_1$ as dictated by  (\ref{check1}-\ref{check2}). The curves start at the attractor value of the scalar, but for any non-zero (positive) perturbation they go to asymptotically flat space (we don't plot the metric functions here, but we have checked this) with a finite value for $\phi_\infty$. As we keep increasing $\phi_1$, from the form of the perturbation in (\ref{horpert1}), we see that at $\phi_1=\sqrt{3}/4 \mu$ the fluctuation from the attractor value of the scalar is zero to leading order in $r$. This suggests that this should correspond to the hairless flat space black hole, and this is indeed what we find from the plots: the horizontal straight line in the plot corresponds to the scalar being a constant equal to the attractor value when $\phi_1=\sqrt{3}/4 \mu$.  

What happens when $\phi_1$ is negative? Even though we don't plot them here because we haven't studied them systematically, numerically it is possible to see that the solutions actually diverge to $-\infty$ at finite radius when $\phi_1$ is negative. The subttractor is a boundary between solutions of this kind, and solutions that go to asymptotically flat spacetimes. It is in this sense that it is a boundary of the attraction basin.  Another immediate question is: what happens when $\phi_\infty$ is positive? We have only considered the lower half of the attraction basin here. It turns out that it is possible to understand these questions analytically in terms of various competing behaviors between $d_1$ and $d_2$ in the exact solution. But since the phase structure of general attractors seems to be quite rich, and since some of these behaviors are more appropriately studied in the context of a more general subtracted geometry, we will defer them for a future occasion \cite{K, AAC}. Some of these subtracted geometries require a different replacement of the warp factor than what has been considered in the literature and we will report on them elsewhere \cite{K}. 

\section*{Acknowledgments}

We would like to thank Justin David, Apoorva Patel and Aninda Sinha for comments and discussions and Avinash Raju for collaboration on related work. CK thanks AC for pressuring/dragging/coaxing him out of his year-long vacation from physics.


\appendix

\section{Connection to ${\cal N}=2$ Supergravity with Four Vector Multiplets}
\label{sugra}

We can relate our action (\ref{minaction}) to the bosonic sector of four dimensional ${\cal N}=2$ supergravity coupled to four vector multiplets. This action has six scalars and four vectors, but we can consistently set three of the scalars to zero, if we are looking at static black hole solutions where a certain pair of gauge field strengths are chosen to be magnetic and the other two are electric \cite{CG}. This reduced action takes the form
\bea
\int d^4x \Big[R- \frac{1}{2} (\partial \varphi_i)^2-\frac{1}{4}\Big(e^{-\varphi_1+\varphi_2-\varphi_3}( F^{1}_{\mu\nu})^2+e^{-\varphi_1+\varphi_2+\varphi_3}(F^{2}_{\mu\nu})^2+\nonumber \hspace{0.5in}\\ 
\hspace{0.5in}+e^{-\varphi_1-\varphi_2+\varphi_3}({\cal F}^{1}_{\mu\nu})^2+e^{-\varphi_1-\varphi_2-\varphi_3}({\cal F}^{2}_{\mu\nu})^2\Big)\Big]
\eea 
where $i=1,2,3$ for $\varphi_i$. Exact static black hole solutions of this action \cite{Chong, CG} can be found in the form (\ref{metric}) and they are described by:
\bea \label{CGBH}
X=r^2-2m r,\ \ \ \Delta_0= \Pi_{I=1}^{4} (r+2m \sinh^2 \delta_I)
\eea 
where all the $\delta_I$ can in principle be different. The scalars take the form
\bea
e^{\varphi_1} = \left[\frac{(r+2m\sinh^2\delta_1)(r+2m\sinh^2\delta_3)}{(r+2m\sinh^2\delta_2)(r+2m\sinh^2\delta_4)}\right]^\frac{1}{2}, \hspace{1.5in} \\
e^{\varphi_2} = \left[\frac{(r+2m\sinh^2\delta_2)(r+2m\sinh^2\delta_3)}{(r+2m\sinh^2\delta_1)(r+2m\sinh^2\delta_4)}\right]^{\frac{1}{2}},
e^{\varphi_3} = \left[\frac{(r+2m\sinh^2\delta_1)(r+2m\sinh^2\delta_2)}{(r+2m\sinh^2\delta_3)(r+2m\sinh^2\delta_4)}\right]^{\frac{1}{2}}.\nonumber
\eea
The gauge fields are of the form
\bea
A_1&=&-2m \cosh \delta_1 \sinh \delta_1 \cos \theta d\phi \\
A_2&=& \frac{2m\cosh \delta_2\sinh \delta_2}{r+2m \sinh^2 \delta_2}  dt \\
{\cal A}_1&=&-2m \cosh \delta_3 \sinh \delta_3 \cos \theta d\phi \\
{\cal A}_2&=&\frac{2m\cosh \delta_4 \sinh \delta_4}{r+2m \sinh^2 \delta_4} dt 
\eea

Computing the field strengths and defining $\varphi_i=2 \phi_i$ we can show that these solutions fall into the attractor ansatz of \cite{Goldstein} with precisely the same normalizations. The charges that are turned on are two magnetic and two electric charges given by:
\bea
q^1_m\equiv 4 Q^0_1=m\sinh 2 \delta_1, \ \ q^2_e\equiv Q^0_2=\frac{m\ \sinh 2 \delta_2}{4}, \\
q^3_m\equiv 4 Q^0_3=m\sinh 2 \delta_3, \ \ q^4_e\equiv  Q^0_4=\frac{m\ \sinh 2 \delta_4}{4}
\eea
The factors of 4 are a consequence of the normalizations in \cite{Goldstein}, which we follow. The $Q^0$'s that are introduced make certain formulas more symmetric, in particular in the extremal limit. With 
\bea
a^2(r)=\frac{X}{\sqrt{\Delta_0}}, \ \ {\rm and} \ \ b^2(r)=\sqrt{\Delta_0},
\eea
the following equations of motion are satisfied:
\bea
(a^2\ b^2)''-2 =0 \\
\frac{b''}{b}+{\phi_i '}^2=0 \\
(a^2 b^2\phi_i')'-\frac{\partial_{\phi_i} V_{eff}(\phi_i)}{2 b^2}=0
\eea
with the energy constraint 
\bea 
a^2{b'}^2+{a^2}'{b^2}'+\frac{V_{eff}(\phi_i)}{b^2}-a^2b^2 {\phi_i'}^2-1=0,
\eea
where 
\bea
V_{eff}(\phi_i)=\frac{1}{4}\Big(m^2\sinh^2 (2 \delta_1)e^{2(-\phi_1+\phi_2-\phi_3)}+m^2\sinh^2 (2 \delta_3)e^{2(-\phi_1-\phi_2+\phi_3)}+\hspace{0.5in}\nonumber\\
\hspace{0.1in}+m^2\sinh^2 (2 \delta_2)e^{2(+\phi_1-\phi_2-\phi_3)}+m^2\sinh^2 (2 \delta_4)e^{2(+\phi_1+\phi_2+\phi_3)}\Big).
\eea

 At this stage it is easy to notice that this set of equations of motion reduce to our minimalistic system (\ref{minansatze1}-\ref{minansatze2}) from section \ref{SubEx}-\ref{AttractAnsatz}, if we set all three $\phi_i\equiv \phi/\sqrt{3}$. The effective potential reduces to (\ref{mineff}) when we work with an action of the form (\ref{minaction}) so that there are only two distinct gauge fields, both of which source magnetic field strengths. Their charges are given by (\ref{magcharge}). This is the way we connect our minimal action (\ref{minaction}) with ${\cal N}=2$ supergravity with four vector multiplets. 

Ultimately the reason why our action is integrable, can be traced to the fact that the original supergravity system is exactly integrable. This is the origin of the diagonalizability/integrability recently found in \cite{JdBoer} in the STU model. We show this and some generalizations by exploiting a known connection with Toda equations in \cite{AAC}.

In the extremal limit of these spacetimes, we have to let 
\bea
m\rightarrow 0, \ \ \ \ \delta_I \rightarrow \infty,
\eea
so that $\frac{m \sinh 2 \delta_I}{4}$ stays finite. The mass formula for the (\ref{CGBH}) black holes takes the form 
\bea
M=\frac{m}{4}\sum_{I=1}^{4} \cosh 2 \delta_I  
\eea
and so in the extremal limit, we have the relation 
\bea
M=Q^0_1+Q^0_2+Q^0_3+Q^0_4.
\eea
The simplicity of this expression was our motivation for introducing the $Q^0_i$.

The subtracted geometries corresponding to (\ref{CGBH}) can be constructed via the warp factor replacement:
\bea
\Delta_s= (2m)^3 r (\Pi_c^2-\Pi_s^2) + (2m)^4 \Pi^2_s.
\eea
This is identical to our prescription (\ref{subrep}) except now the charges are different, so 
 \bea
\Pi_c=\Pi_{I=1}^4\cosh \delta_I, \ \ \ \Pi_s =\Pi_{I=1}^4\sinh \delta_I.
\eea
The subtracted geometry takes a simple form in the extremal limit (and we adapted this form to construct our minimalist solution in Section \ref{SubEx}), and is again defined by
\bea
X= r^2, \ \ \ \Delta_s=\frac{\alpha^4}{\mu} (r +\mu)
\eea
but now with
\bea \label{subt1charge}
\Big(\frac{\alpha}{4}\Big)^4\equiv \Pi_{I=1}^4 Q^0_I, \ \ \ \ \frac{4}{\mu}\equiv\sum_{I=1}^4 \frac{1}{Q^0_I}.
\eea

\section{Conical Box Asymptotics}
\label{Box}

Since the change in the asymptotics is a crucial point of the paper, in this appendix we discuss some aspects of the boundary structure of the subtracted geometry. First, we write the metric defined by (\ref{subrep}-\ref{pipi}) in a form that manifests its asymptotic structure by defining a better radial coordinate, $R^2 \equiv\sqrt{\Delta_s}$. In terms of this coordinate, the subtracted geometry becomes
\bea
ds^2=-\frac{(R^4-q-mp)^2-m^2p^2}{p^2 \ R^2} dt^2+\frac{16 R^8}{(R^4-q-mp)^2-m^2p^2}dR^2+R^2 d\Omega_2^2,
\eea
where 
\bea
p\equiv(2m)^3 (\Pi_c^2-\Pi_s^2),\ \ \ q\equiv(2m)^4 \Pi_s^2
\eea
It can be seen that in the extremal limit (where $m \rightarrow 0$ with $q=\alpha^4, p=\alpha^4/\mu$ staying finite) this change of coordinates results in $\mu$ ending up as a pure rescaling of the time coordinate. 
In any event, the radial coordinate has a straightforward interpretation now, and asymptotically the metric looks like
\bea\label{conBox}
ds^2_{\rm box}=-\frac{R^6}{p^2} dt^2+16 dR^2 +R^2 d\Omega_2^2.
\eea
This is called the conical box \cite{CG} and this is where all our subtracted black holes live. Note that the charged/extremal/Schwarzschild/... black holes all live in this universal box. 

It is always tempting to put flat space black holes in boxes. This is because in flat space there are conceptual difficulties associated with defining Hartle-Hawking states etc. \cite{CKreview}, and one hopes that the thermodynamics of flat space black holes is a real-world/local-equilibrium reflection of a Platonic boxed black hole. So as a first step, one might try to consider the dynamics of the empty box by itself and perhaps try to do quantum field theory in this background, explore an analogue of Hawking-Page transition, etc. similar to the AdS-box. But this leads to immediate technical and conceptual difficulties. One big problem is that it is not clear that one should think of the empty box as a ground state, because we need matter to support the subtracted geometry. A well-known manifestation of this problem is that the empty box has a naked curvature singularity at the origin\footnote{We thank Aninda Sinha for emphasizing this to us.}. We suspect that one way forward might be to consider the extremal black hole as the vacuum state instead of the empty box \cite{new}. 

We conclude this appendix by drawing the Penrose diagram of the conical box (\ref{conBox}). As usual, after suppressing the sphere part and extracting a conformal factor the relevant piece of the metric takes the form
\bea
d\hat s^2=-dt^2+\frac{16 p^2}{R^6}dR^2= -dt^2+dz^2 =-du \ dv.
\eea 
We have defined 
\bea
z=-\frac{2p}{R^2}, \ \ \ u\equiv t-z,\ \ v\equiv t+z.
\eea
Note that with this definition $z$ is increasing when $R$ is increasing. $R$ is in the range $(0,\infty)$ and $t$ is in the range $(-\infty, \infty)$. So $z \in (-\infty,0^-)$ and $u,v \in (-\infty, \infty)$. At this stage we can compactify the manifold by defining $u=\tan(\tilde u/2), \ v=\tan(\tilde v/2)$, extract a confromal factor and we are left with the flat space metric but with a finite range for the coordinates: $\tilde u, \tilde v \in (-\pi, \pi)$. This means that the Penrose diagram is at most a square (diamond). It need not span the entire square because $R$ is restricted to be real (and positive), whereas the final coordinate change treats $R^2$ as an arbitrary real number.

\begin{figure}
\begin{center}
\includegraphics[height=0.70\textheight]{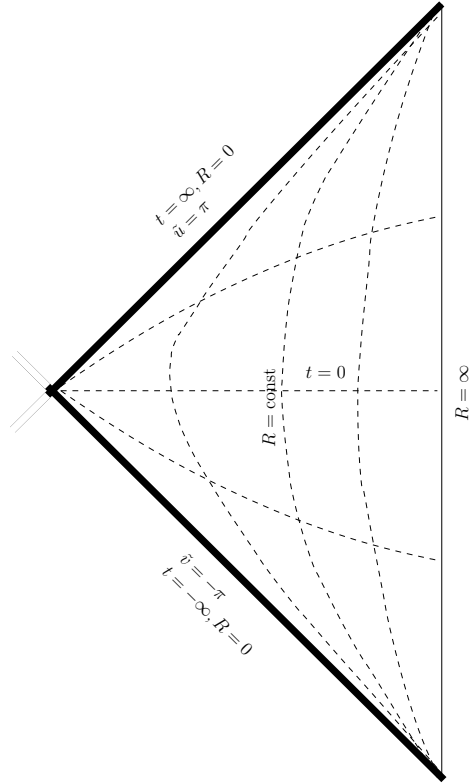}
\caption{Penrose diagram of (\ref{conBox}). The bold line segments indicate the curvature singularities at $R=0$, the boundary at $R=\infty$ is marked. The spacetime can be foliated with constant $R$ curves (\ref{curves1}) or constant $t$ curves (\ref{curves2}) which are indicated schematically in the figure. The natural comparison of this figure is against the Poincare patch of AdS, with the Poincare horizons replaced by the curvature singularity here.}
\end{center}
\end{figure}

To get some intuition about the geometry, we can plot 
\bea\label{curves1}
\frac{4 p}{R^2} = \tan \frac{\tilde u}{2}-\tan \frac{\tilde v}{2}, \\
2 t=\tan \frac{\tilde u}{2}+\tan \frac{\tilde v}{2}\label{curves2}
\eea
in the $\tilde u, \tilde v$ square for various values of $t$ and $R^2$, specifically for $t, R^2 \rightarrow 0$ or $\pm\infty$. The region spanned by negative values of $R^2$ have to be omitted, which results in the Penrose diagram becoming a triangle instead of a square diamond. The end result is given in figure 2. It is clear that the causal structure and $R \rightarrow \infty$ asymptotics are very different from flat space, and indeed is very similar to the Poincare patch of AdS, except now instead of the Poincare horizon, we have a curvature singularity. This similarity is perhaps not surprising considering the fact that we are working with a box here.

When we put a black hole in, we expect that the interior (the region close to $r=0$, the inner/outer horizons) of the Penrose diagram will change and will be similar to the corresponding flat space black holes, but the boundary will remain a vertical line. For charged black holes there will be many asymptotic regions and the diagram will be infinite, just as in flat space and AdS cases, but we do not expect any substantial qualitative changes in the general structure of the diagram. Similar statements should hold also for extremal black holes where the two horizons collapse onto each other.

%

\begin{thebibliography}{19}        


\bibitem{StromingerVafa} 
  A.~Strominger and C.~Vafa,
  ``Microscopic origin of the Bekenstein-Hawking entropy,''
  Phys.\ Lett.\ B {\bf 379}, 99 (1996)
  [hep-th/9601029].

\bibitem{FKS} 
  S.~Ferrara, R.~Kallosh and A.~Strominger,
  ``N=2 extremal black holes,''
  Phys.\ Rev.\ D {\bf 52}, 5412 (1995)
  [hep-th/9508072].

\bibitem{SenReview} 
  A.~Sen,
  ``Black Hole Entropy Function, Attractors and Precision Counting of Microstates,''
  Gen.\ Rel.\ Grav.\  {\bf 40}, 2249 (2008)
  [arXiv:0708.1270 [hep-th]].

\bibitem{SenException} 
  N.~Banerjee, I.~Mandal and A.~Sen,
  ``Black Hole Hair Removal,''
  JHEP {\bf 0907}, 091 (2009)
  [arXiv:0901.0359 [hep-th]].

\bibitem{Maldacena}
  J.~M.~Maldacena,
  ``The large N limit of superconformal field theories and supergravity,''
  Adv.\ Theor.\ Math.\ Phys.\  {\bf 2}, 231 (1998)
  [Int.\ J.\ Theor.\ Phys.\  {\bf 38}, 1113 (1999)]
  [arXiv:hep-th/9711200].


\bibitem{tHooft} 
  G.~'t Hooft,
  ``Dimensional reduction in quantum gravity,''
  gr-qc/9310026.
\bibitem{Susskind} 
  L.~Susskind,
  ``The World as a hologram,''
  J.\ Math.\ Phys.\  {\bf 36}, 6377 (1995)
  [hep-th/9409089].

\bibitem{DabholkarSen} 
  A.~Dabholkar, A.~Sen and S.~P.~Trivedi,
  ``Black hole microstates and attractor without supersymmetry,''
  JHEP {\bf 0701}, 096 (2007)
  [hep-th/0611143].

\bibitem{Goldstein} 
  K.~Goldstein, N.~Iizuka, R.~P.~Jena and S.~P.~Trivedi,
  ``Non-supersymmetric attractors,''
  Phys.\ Rev.\ D {\bf 72}, 124021 (2005)
  [hep-th/0507096].

\bibitem{CveticLarsenOld} 
  M.~Cvetic and F.~Larsen,
  ``General rotating black holes in string theory: Grey body factors and event horizons,''
  Phys.\ Rev.\ D {\bf 56}, 4994 (1997)
  [hep-th/9705192].

\bibitem{StromKerrCFT} 
  M.~Guica, T.~Hartman, W.~Song and A.~Strominger,
  ``The Kerr/CFT Correspondence,''
  Phys.\ Rev.\ D {\bf 80}, 124008 (2009)
  [arXiv:0809.4266 [hep-th]].

\bibitem{LMP} 
  H.~Lu, J.~Mei and C.~N.~Pope,
  ``Kerr/CFT Correspondence in Diverse Dimensions,''
  JHEP {\bf 0904}, 054 (2009)
  [arXiv:0811.2225 [hep-th]].
  
\bibitem{Hartman} 
  T.~Hartman, K.~Murata, T.~Nishioka and A.~Strominger,
  ``CFT Duals for Extreme Black Holes,''
  JHEP {\bf 0904}, 019 (2009)
  [arXiv:0811.4393 [hep-th]].

\bibitem{Stanislav} 
  C.~Krishnan and S.~Kuperstein,
  ``A Comment on Kerr-CFT and Wald Entropy,''
  Phys.\ Lett.\ B {\bf 677}, 326 (2009)
  [arXiv:0903.2169 [hep-th]].

\bibitem{Geoffrey} 
  G.~Compere,
  ``The Kerr/CFT correspondence and its extensions: a comprehensive review,''
  Living Rev.\ Rel.\  {\bf 15}, 11 (2012)
  [arXiv:1203.3561 [hep-th]].

\bibitem{HiddenCS} 
  A.~Castro, A.~Maloney and A.~Strominger,
  ``Hidden Conformal Symmetry of the Kerr Black Hole,''
  Phys.\ Rev.\ D {\bf 82}, 024008 (2010)
  [arXiv:1004.0996 [hep-th]].


\bibitem{CK} 
  C.~Krishnan,
  ``Hidden Conformal Symmetries of Five-Dimensional Black Holes,''
  JHEP {\bf 1007}, 039 (2010)
  [arXiv:1004.3537 [hep-th]].

\bibitem{CL1} 
  M.~Cvetic and F.~Larsen,
  ``Conformal Symmetry for General Black Holes,''
  JHEP {\bf 1202}, 122 (2012)
  [arXiv:1106.3341 [hep-th]].


\bibitem{CL2} 
  M.~Cvetic and F.~Larsen,
  ``Conformal Symmetry for Black Holes in Four Dimensions,''
  JHEP {\bf 1209}, 076 (2012)
  [arXiv:1112.4846 [hep-th]].

\bibitem{CG} 
  M.~Cvetic and G.~W.~Gibbons,
  ``Conformal Symmetry of a Black Hole as a Scaling Limit: A Black Hole in an Asymptotically Conical Box,''
  JHEP {\bf 1207}, 014 (2012)
  [arXiv:1201.0601 [hep-th]].

\bibitem{Ami} 
  A.~Virmani,
  ``Subtracted Geometry From Harrison Transformations,''
  JHEP {\bf 1207}, 086 (2012)
  [arXiv:1203.5088 [hep-th]].

\bibitem{GibMaed} 
  G.~W.~Gibbons and K.~-i.~Maeda,
  ``Black Holes and Membranes in Higher Dimensional Theories with Dilaton Fields,''
  Nucl.\ Phys.\ B {\bf 298}, 741 (1988).

\bibitem{Pope} 
  H.~Lu and C.~N.~Pope,
  ``SL(N+1,R) Toda solitons in supergravities,''
  Int.\ J.\ Mod.\ Phys.\ A {\bf 12}, 2061 (1997)
  [hep-th/9607027].
  
\bibitem{K}
``Generalized Subtracted Geometries and the Phases of Attraction", to appear.

\bibitem{CveticYoum} 
  M.~Cvetic and D.~Youm,
  ``General rotating five-dimensional black holes of toroidally compactified heterotic string,''
  Nucl.\ Phys.\ B {\bf 476}, 118 (1996)
  [hep-th/9603100].
 
\bibitem{AAC}
A. Chakraborty, C. Krishnan and A. Raju, ``Attraction, Subtraction and Toda  Solutions: STU model and Beyond", to appear.

\bibitem{Chong} 
  Z.~-W.~Chong, M.~Cvetic, H.~Lu and C.~N.~Pope,
  ``Charged rotating black holes in four-dimensional gauged and ungauged supergravities,''
  Nucl.\ Phys.\ B {\bf 717}, 246 (2005)
  [hep-th/0411045].

\bibitem{JdBoer} 
  M.~Baggio, J.~de Boer, J.~I.~Jottar and D.~R.~Mayerson,
  ``Conformal Symmetry for Black Holes in Four Dimensions and Irrelevant Deformations,''
  arXiv:1210.7695 [hep-th].

\bibitem{CKreview} 
  C.~Krishnan,
  ``Quantum Field Theory, Black Holes and Holography,''
  arXiv:1011.5875 [hep-th].

\bibitem{new} Work in progress.


\end{thebibliography}

\end{document}